\documentclass[12pt,a4paper]{article}
\usepackage[utf8]{inputenc}
\usepackage{amsmath}
\usepackage{amsfonts}
\usepackage{amssymb}
\usepackage{url}
\usepackage{natbib}
\usepackage{tabularx}
\usepackage{float}
\usepackage{booktabs}
\usepackage{lineno}
\usepackage{changepage}

\usepackage{graphicx}
\usepackage[left=2cm,right=2cm,top=2cm,bottom=2cm]{geometry}

\title{Prediction of motor insurance claims occurrence as an imbalanced machine learning problem}

\author{Sebastian Baran\\
Department of Mathematics\\ 
Cracow University of Economics\\
31-510 Cracow, Poland\\
E-mail: sebastian.baran@uek.krakow.pl
\and 
Przemys\l aw Rola\\
Department of Mathematics\\ 
Cracow University of Economics\\
31-510 Cracow, Poland\\
E-mail: przemyslaw.rola@uek.krakow.pl}

\date{}

\begin{document}

\maketitle

\abstract{The insurance industry, with its large datasets, is a natural place to use big data solutions. However it must be stressed, that significant number of applications for machine learning in insurance industry, like fraud detection or claim prediction, deals with the problem of machine learning on an imbalanced data set. This is due to the fact that frauds or claims are rare events when compared with the entire population of drivers. The problem of imbalanced learning is often hard to overcome. Therefore, the main goal of this work is to present and apply various methods of dealing with an imbalanced dataset in the context of claim occurrence prediction in car insurance. In addition, the above techniques are used to compare the results of machine learning algorithms in the context of claim occurrence prediction in car insurance. Our study covers the following techniques: logistic-regression, decision tree, random forest, xgBoost, feed-forward network. The problem is the classification one.}

\section{Introduction}
At the moment, we have a time of universal digitization. Our phones, computers, online shopping, social networks are tools that enable the creation of huge data sets. Huge databases, the rapid development of computing power and the availability of technologically advanced solutions such as GPU computing for an ordinary home computer user contributed to the dynamic development of machine learning.

There is a significant number of applications for machine learning in industries available, including product recommendations for online shopping, fraud detection in banks, improving customer service in retail or even drug discovery in healthcare. Since insurance is a traditional business with many regulations, big players were initially reluctant to deviate from an approach that was tried and tested through the ages. However, in recent years, we can observe a dynamic development of machine learning applications in the insurance industry. These applications include but are not limited to virtual assistants for advising potential customers, determination of risk profiles for underwriting, fraud detection and prevention, claims processing and prediction and customer retention. Another confirmation of the increase in the use of machine learning methods in insurance is the growing number of jobs offers for data scientists and machine learning engineers in insurance industries.

It must be emphasized here that the insurance industry, during its many years of activity, had the opportunity to collect a huge amount of data. If, for example, we consider car insurance, each insurance policy contains a lot of information about the customer, car and place of use of the car. If we add telematics data to this, the datasets become really large and the use of machine learning methods to process the information contained in these sets seems to be a natural solution.

There are two groups in the insurance industry: life insurance and non-life insurance. This paper considers non-life insurance, particularly auto insurance. Insurance claims occur when the policyholder (the customer) creates a formal request to an insurer for coverage or compensation for an accident. The insurance company must validate this request and then decide whether to issue payment to the policyholder. There are many factors that determine the price of car insurance. The driver's skills, the technical condition of the car, the population density in the place where the car is used are the first factors that come to mind. The number of claims caused by the driver in previous years is also very important. If a client has a good driving record, it would be unreasonable for a client with a poor driving background to pay a similar insurance premium. Hence the popularity of the so-called bonus-malus system in insurance. A bonus-malus system (BMS) is a system that adjusts the premium paid by a customer according to their individual claim history. Bonus usually is a discount in the premium which is given on the renewal of the policy if no claim is made in the previous year. Malus is an increase in the premium if there is a claim in the previous year. Bonus-malus systems are very common in vehicle insurance. This system is also called a no-claim discount (NCD) or no-claims bonus in Britain and Australia. Insurance companies want to have an insurance premium that is appropriate for each customer. Insurance companies are looking for an insurance premium that is appropriate for each client. They are looking for a model that will allow them to react any changing circumstances and control any losses. 

It is worth noting that the COVID-19 pandemic has a significant impact on the current situation on the car insurance market. According to the Insurance Information Institute report\footnote{\url{https://www.iii.org/sites/default/files/docs/pdf/triple-i_auto_insurance_rates_02112022.pdf}} from the beginning of 2022 auto insurance premiums have returned to pre-pandemic levels after \$14 billion was returned to policyholders in anticipation of fewer accidents. But something interesting happened during the pandemic. While miles driven declined and accident frequency initially dropped, frequency and severity quickly started increasing again. Collision frequency and severity for US car insurance increased by 42\% and 43\% respectively (year-over-year, 3Q 2021). These factors combined with replacement parts inflation due to supply-chain issues have driven insurers auto losses above pre-pandemic levels. The factors outlined above will continue to put pressure on insurers to raise rates to ensure the coverage is appropriately priced. In order to follow the growing trends in claim severity and frequency an efficient system for filing auto insurance claims is needed. These facts make auto insurance pricing studies more meaningful and essential.

It has been stated above that the insurance industry, with its large datasets, is a natural place to use big data solutions. However it must be stressed, that significant number of applications for machine learning in insurance industry, like fraud detection or claim prediction, deals with the problem of machine learning on an imbalanced data set. This is due to the fact that frauds or claims are rare events when compared with the entire population of drivers. The problem of imbalanced learning is often hard to overcome. Therefore, the main goal of this work is to present and apply various methods of dealing with an imbalanced dataset in the context of claim occurrence prediction in car insurance. In addition, the above techniques are used to compare the results of machine learning algorithms in the context of claim occurrence prediction in car insurance. Our study covers the following techniques: logistic-regression, decision tree, random forest, xgBoost, feed-forward network.

\section{Related Work}
A comprehensive summary of the papers in which claim prediction was discussed, both in terms of frequency and severity, can be found in 1. In this chapter, we focus on the studies that considered the problem of claim occurrence. \citeauthor{Smith} (\citeyear{Smith}) tested several machine learning models, like the decision tree and neural networks, to assess whether the policyholder submits a claim or not and addressed the effect that the case study will have on the insurance company. This study shows that the neural network model is better than decision tree. \citeauthor{Jing} (\citeyear{Jing}) used only a Bayesian network to classify either a claim or no claim. \citeauthor{Pesantez} (\citeyear{Pesantez}) use two competing methods, XGBoost and logistic regression, to predict the frequency of motor insurance claims. This study shows that the XGBoost model is slightly better than logistic regression; however, they used a database comprised of only 2767 observations. Furthermore, a model for predicting insurance claims was developed (\citeauthor{Abdelhadi} \citeyear{Abdelhadi}); they built four classifiers to predict the claims occurance, including XGBoost, J48, ANN, and na?ve Bayes algorithms. The XGBoost model performed the best among the four models, and they used a database comprised of 30,240 observations.

All of the above studies considered neither big volume nor missing value issues. Moreover most of them considered only few machine learning algorithms. \citeauthor{Hanafy} (\citeyear{Hanafy}) used big data that contained almost a million and a half (1,488,028) observations with 59 variables. This study shows that XGBoost is a useful model. However random forest and the decision tree (C50) is significantly better than XGBoost. and the na?ve Bayes is the worst model for predicting claims occurrence among all eight classification models used in this study. Due to the very large dataset and the large number of machine learning algorithms used, this work seems to be very promising. Especially because we are dealing here with a real database provided by Porto Seguro\footnote{\url{https://www.kaggle.com/alinecristini/atividade2portoseguro}} company. However, there is one point that prompts us to treat the results of the above paper with great caution. The authors of this article, during dataset preprocessing, used ROSE algorithm in order to overcome data imbalance. Balancing dataset it is standard and correct procedure when we deal with imbalanced learning problem. However, authors applied this algorithm to the full datasetset, whilst any procedure that balance dataset like ROSE or SMOTE should be used only on training data. Balancing algorithm like ROSE is used on training data to ensure greater representation of the minority class. However test data should not be balanced, because the test data is the equivalent of real imbalanced data and on these imbalanced data machine learning algorithm is supposed to work well. Since Hanafy and Ming train the algorithm on a balanced dataset and also test on a balanced dataset (which in fact does not happen when we get real data from policies), hence the good results from this article are no longer surprising. Moreover, Porto Seguro dataset and the claim occurrence task was the subject of the Kaggle competition with a prize of 25,000 dollars\footnote{\url{https://www.kaggle.com/c/porto-seguro-safe-driver-prediction/overview/description}}. The winning solution by Michael Jahrer\footnote{\url{https://www.kaggle.com/c/porto-seguro-safe-driver-prediction/discussion/44629}} was a blend of one LightGBM and five neural networks trained on denoising autoencoder hidden activations, in order to learn a better representation of the numeric data. This advanced model has obtained score 0.29698 (Normalized Gini Coefficient (NGI)). Since NGI = 2*AUC - 1, hence the winning score is equivalent to area under curve AUC = 0.64849. Using basic machine learning algorithms and adding only hyperparameter tuning to it, Hanafy and Ming obtained much better results than the winning solution from the competition (AUC = 0.84 for random forest, AUC=0.769 for C50). However, these better results, as mentioned earlier, are due to incorrectly applied balancing procedure. It is worth emphasizing here that article by \citeauthor{Hanafy} (\citeyear{Hanafy}), despite the issue with balancing procedure, is very interesting, valuable, well-organized and it was a motivation to write this paper.

% In ~\citeauthor{Hanafy}, the authors, investigated some ML techniques to predict claims occurrence by analyzing the big dataset given by Porto Seguro\footnote{https://www.kaggle.com/alinecristini/atividade2portoseguro}, a large automotive company based in Brazil. We performed similar analysis on a new dataset Fremotor1 described in the next section.

\section{Background}
\subsection{Imbalanced classification}
Car insurance claims are an excellent example of imbalanced data, because the majority of policyholders do not make a claim. Therefore, to understand the problem, it is essential to have a knowledge about imbalanced classification. We explore this term following \citeauthor{Fernandez} (\citeyear{Fernandez}). 

Generally, any dataset with an unequal class distribution is technically imbalanced. However, a dataset is said to be imbalanced when there is a significant, or in some cases extreme, disproportion among the number of examples of each class of the problem. Most of the imbalanced classification literature considers binary classification problems, where one class significantly outnumbers the other. In two-class problems the minority (underrepresented) class is usually referred to as the positive class, whereas the majority class is considered to be the negative one. These terms are used interchangeably in the literature.

Obviously, one would want a 100\% of accuracy for both classes. However, classifiers are usually far from being perfect and they tend to have a great accuracy for the majority class while obtaining poor results (closer to 0\%) for the minority class. Unfortunately, standard classifier learning algorithms are usually biased toward the majority class, since rules correctly predicting those instances are positively weighted in favour of the accuracy metric or the corresponding cost function.

Accuracy is not a proper metric in the case of imbalanced dataset, since it does not distinguish between the numbers of correctly classified examples of different classes. We may even encounter the so-called accuracy paradox. For example, consider a dataset with a ratio of 1:100 between positive and negative class and an algorithm which classifies all examples as negatives. On the one hand, such a classifier achieved an impressive 99\% accuracy. On the other hand, it is completely useless in detecting positive class instances which is the class of interest of the problem from the application point of view.

\subsection{Performance metrics}

\subsubsection{Confusion matrix}
A confusion matrix is used for binary classification problems. It is a beneficial method to distinguish which class outputs were predicted correctly or not. In the matrix, TP and TN represent the quantity of correctly classified positive and negative instances, whereas FP and FN represent incorrectly classified positive and negative samples, respectively.

\begin{table}[H] 
\caption{Confusion matrix.\label{conf_matrix}}
\newcolumntype{C}{>{\centering\arraybackslash}X}
\begin{tabularx}{\textwidth}{CCC}
\toprule
 & \textbf{Predicted Positive} & \textbf{Predicted Negative}\\
\midrule
\textbf{Actual Positive} & True positive (TP) &  False negative(FN) \\ 
\textbf{Actual Negative} & False positive (FP) &  True negative (TN) \\ 
\bottomrule
\end{tabularx}
\end{table}
\unskip

\subsubsection{Precision and Recall}

The precision metric is used to measure the positive class that are correctly predicted from the total predicted in a positive class. Recall is used to measure the fraction of positive class that are correctly classified; this basically describes how well the model can detect the class type (\cite{Hossin}).

\begin{linenomath}
\begin{equation}
\textit{Precision} = \frac{TP}{TP + FP},
\end{equation}
\end{linenomath}

\begin{linenomath}
\begin{equation}
\textit{Recall} = \frac{TP}{TP + FN},
\end{equation}
\end{linenomath}

If we want to seek for the balance between precision and recall there is another common measure. $\textit{F1}$ is the harmonic mean of precision and recall:
\begin{linenomath}
\begin{equation}
\textit{F1} = \frac{2 \cdot \textit{Precision} \cdot \textit{Recall}}{\textit{Precision} + \textit{Recall}}.
\end{equation}
\end{linenomath}

\section{Materials and Methods}

Even a good ML algorithm might not perform well for imbalanced data. To overcome this problem, we applied the oversampling technique, i.e. generating more representations of positive class, so the data becomes more balanced. We used SMOTE oversampling technique, i.e. Synthetic Minority Over-sampling Technique.

The basic idea behind the SMOTE (in case of continuous variable) is:
\begin{itemize}
    \item random sample a point $q$ from the minority class
    \item specify $k$ (e.g. $k=5$) nearest neighbors (kNN)
    \item randomly choose one point among them ($x_i$)
    \item the synthetic point is then sampled from the interval created by $q$ and $x_i$ in the feature space.
\end{itemize}

\subsection{Dataset}

Fremotor1 datset is a part of the R package - CASdatasets, previously available with the book \textit{Computational Actuarial Science with R} of Arthura Charpentiera.
It consists of  9 subsets fremotor1freq0304a/b/c, fremotor1sev0304a/b/c, fremotor1prem0304a/b/c describing the claims of car insurance and parameters of the insurance policy of the one unknown French insurer from 2003 - 2004.
\begin{itemize}
    \item datasets fremotor1freq0304a/b/c contain 64.234 records describing the number of claims from various guarantees for policy from 2003 - 2004;
    \item datasets fremotor1prem0304a/b/c contain 51.949 records with explanatory variables of policy from 2003 - 2004 (very likely with multiple vehicles insured with the same policy number); 
\item datasets fremotor1sev0304a/b/c
contain 9.246 records with the amount of claim, dates of the claim, appropriate guarantee, from 2003 and 2004.
\end{itemize}

\begin{table}[H] 
\caption{Dataset.\label{tab_dataset}}
\newcolumntype{C}{>{\centering\arraybackslash}X}
\begin{tabularx}{\textwidth}{CCCC}
\toprule
  & \textbf{dtype} & \textbf{num\_missing} & \textbf{num\_uniques} \\
\midrule
\textbf{NbClaimsTot}   & int64          & 0                     & 2                     \\
\textbf{DrivAge}       & int64          & 0                     & 72                    \\
\textbf{DrivGender}    & object         & 0                     & 2                     \\
\textbf{MaritalStatus} & object         & 34795                 & 5                     \\
\textbf{BonusMalus}    & int64          & 0                     & 68                    \\
\textbf{LicenceNb}     & int64          & 0                     & 7                     \\
\textbf{PayFreq}       & object         & 0                     & 4                     \\
\textbf{JobCode}       & object         & 34795                 & 7                     \\
\textbf{VehAge}        & int64          & 0                     & 49                    \\
\textbf{VehClass}      & object         & 0                     & 9                     \\
\textbf{VehPower}      & object         & 0                     & 15                    \\
\textbf{VehGas}        & object         & 0                     & 2                     \\
\textbf{VehUsage}      & object         & 0                     & 3                     \\
\textbf{Garage}        & object         & 0                     & 4                     \\
\textbf{Area}          & object         & 0                     & 10                    \\
\textbf{Region}        & object         & 0                     & 4                     \\
\textbf{Channel}       & object         & 0                     & 3                     \\
\textbf{Marketing}     & object         & 0                     & 4
\\
\bottomrule
\end{tabularx}
\end{table}

%Materials and Methods should be described with sufficient details to allow others to replicate and build on published results. Please note that publication of your manuscript implicates that you must make all materials, data, computer code, and protocols associated with the publication available to readers. Please disclose at the submission stage any restrictions on the availability of materials or information. New methods and protocols should be described in detail while well-established methods can be briefly described and appropriately cited.

%Research manuscripts reporting large datasets that are deposited in a publicly avail-able database should specify where the data have been deposited and provide the relevant accession numbers. If the accession numbers have not yet been obtained at the time of submission, please state that they will be provided during review. They must be provided prior to publication.

%Interventionary studies involving animals or humans, and other studies require ethical approval must list the authority that provided approval and the corresponding ethical approval code.
%\begin{quote}
%This is an example of a quote.
%\end{quote}

%%%%%%%%%%%%%%%%%%%%%%%%%%%%%%%%%%%%%%%%%%
\section{Results}

\begin{table}[H] 
\caption{ML results with the default hyperparameters.\label{tab1}}
\newcolumntype{C}{>{\centering\arraybackslash}X}
\begin{tabularx}{\textwidth}{CCCCCCC}
\toprule
 & \textbf{accuracy} & \textbf{F1} & \textbf{precision} & \textbf{recall} & \textbf{AUC} & \textbf{AUPRC}\\
\midrule
\textbf{Logistic Regression} & 0.8652 & 0      & 0      & 0      & 0.5    & 0.163  \\ 
\textbf{Decision Tree}       & 0.8193 & 0.3323 & 0.331  & 0.3336 & 0.6143 & 0.2057 \\ 
\textbf{Random Forest}       & 0.8513 & 0.3416 & 0.4236 & 0.2862 & 0.6128 & 0.2781 \\ 
\textbf{XGBoost}             & 0.8639 & 0.0236 & 0.3542 & 0.0122 & 0.5044 & 0.197  \\ 
\bottomrule
\end{tabularx}
\end{table}
\unskip

\begin{figure}[H]
\caption{Confusion matrix of ML models with the default hyperparameters}
\centering
\includegraphics[scale=0.65]{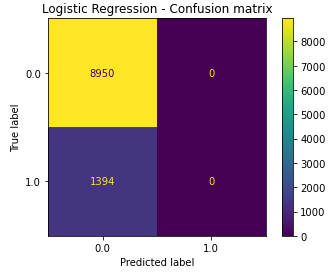}
\includegraphics[scale=0.65]{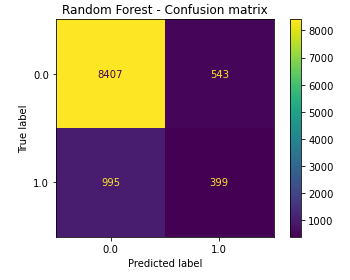}
\centering
\includegraphics[scale=0.65]{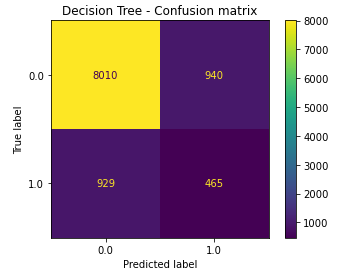}
\includegraphics[scale=0.65]{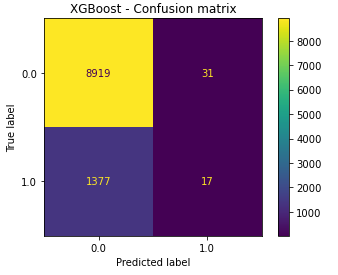}

\end{figure}

\begin{table}[H]
\caption{Hyperparameter tuning after SMOTE - Sklearn GridSearchCV\label{opt_hyp1}}
\resizebox{\textwidth}{!}{%
\begin{tabular}{|c|c|c|}
\hline
 &
  \textbf{Parameters Grid} &
  \textbf{Best Parameters} \\ \hline
\textbf{Logistic Regression} &
  \begin{tabular}[c]{@{}c@{}}max\_iter': {[}20, 50, 100, 200,   500, 1000{]},\\      'solver': {[}'newton-cg', 'lbfgs', 'liblinear', 'sag', 'saga'{]}\end{tabular} &
  \begin{tabular}[c]{@{}c@{}}max\_iter': 50\\      'solver': 'sag'\end{tabular} \\ \hline
\textbf{Decision Tree} &
  \begin{tabular}[c]{@{}c@{}}criterion': {[}'gini', 'entropy'{]},   \\      'max\_depth': {[}5, 15, 25, 35, None{]}, \\      'max\_features': {[}'auto', 'log2', None{]}, \\      'min\_samples\_split': {[}2, 9{]}, \\      'min\_samples\_leaf': {[}1, 8{]}\end{tabular} &
  \begin{tabular}[c]{@{}c@{}}criterion': 'entropy', \\      'max\_depth': None, \\      'max\_features': 'None', \\      'min\_samples\_split': 2, \\      'min\_samples\_leaf': 1\end{tabular} \\ \hline
\textbf{Random Forest} &
  \begin{tabular}[c]{@{}c@{}}criterion': {[}'gini', 'entropy'{]},   \\      'n\_estimators': {[}100, 200, 300{]}, \\      'max\_depth': {[}5, 15, 25, None{]}, \\      'max\_features': {[}'auto', 'log2', None{]}\end{tabular} &
  \begin{tabular}[c]{@{}c@{}}criterion': 'gini', \\      'n\_estimators': 300, \\      'max\_depth': None, \\      'max\_features': 'log2'\end{tabular} \\ \hline
\textbf{XGBoost} &
  \begin{tabular}[c]{@{}c@{}}n\_estimators': {[}100, 200, 300{]},   \\      'max\_depth': {[}6, 12, 18{]}, \\      'learning\_rate': {[}0.1, 0.5, 0.9, 0.95{]}, \\      'subsample': {[}0.6, 1.0{]}\end{tabular} &
  \begin{tabular}[c]{@{}c@{}}'n\_estimators': 300, \\      'max\_depth': 12, \\      'learning\_rate': 0.1, \\      'subsample': 1.0\end{tabular} \\ \hline
\end{tabular}%
}
%\label{tab:my-table}
\end{table}

\begin{table}[H]
\caption{Hyperparameter tuning after SMOTE - Keras Tuner Hyperband\label{opt_hyp2}}
\resizebox{\textwidth}{!}{%
\begin{tabular}{|c|c|c|}
\hline
 &
  \textbf{Parameters Grid} &
  \textbf{Best Parameters} \\ \hline
\textbf{DL} &
  \begin{tabular}[c]{@{}c@{}}batch\_size': {[}128, 256,   512{]}\\      'num\_layers': 1 - 3\\      'units\_1': 16 - 256, step 16\\      'units\_2': 16 - 256, step 16\\      'units\_3': 16 - 256, step 16\\      'learning\_rate': {[}0.1, 0.01, 0.001, 1e-4, 1e-5, 1e-6{]}\end{tabular} &
  \begin{tabular}[c]{@{}c@{}}batch\_size': 256,\\      'num\_layers': 3,\\      'units\_1': 48,\\      'units\_2': 176,\\      'units\_2': 208,\\      'learning\_rate': 0.001\end{tabular} \\ \hline
\end{tabular}%
}

%\label{tab:my-table}
\end{table}

\begin{table}[H] 
\caption{Results after SMOTE and hyperparameter tuning.\label{tab2}}
\newcolumntype{C}{>{\centering\arraybackslash}X}
\begin{tabularx}{\textwidth}{CCCCCCC}
\toprule
 & \textbf{accuracy} & \textbf{F1} & \textbf{precision} & \textbf{recall} & \textbf{AUC} & \textbf{AUPRC}\\
\midrule
\textbf{Logistic Regression} & 0.5236 & 0.2407 & 0.1533 & 0.5603 & 0.5391 & 0.1586 \\ 
\textbf{Decision Tree}       & 0.8164 & 0.3297 & 0.3245 & 0.335  & 0.6132 & 0.2011 \\ 
\textbf{Random Forest}       & 0.8464 & 0.358  & 0.4098 & 0.3178 & 0.6233 & 0.2743 \\ 
\textbf{XGBoost}             & 0.8551 & 0.2148 & 0.3981 & 0.1471 & 0.5562 & 0.2583 \\ 
\textbf{DL}                  & 0.7991 & 0.3200 & 0.2942 & 0.3508 & 0.6385 & 0.2403 \\ 
\bottomrule
\end{tabularx}
\end{table}
\unskip

\begin{figure}[H]
\caption{Confusion matrix of ML models after SMOTE and hyperparameter tuning.\label{conf_matrix_SMOTE}}
%\column{0.33\textwidth}
\centering
\includegraphics[scale=0.65]{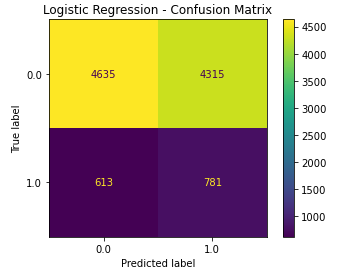}
\includegraphics[scale=0.65]{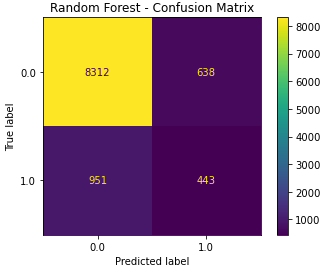}
%\column{0.33\textwidth}
\centering
\includegraphics[scale=0.65]{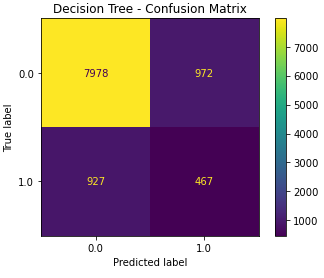}
\includegraphics[scale=0.65]{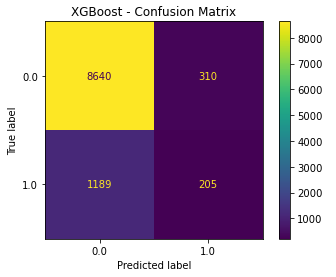}
%\column{0.33\textwidth}
\centering
\includegraphics[scale=0.65]{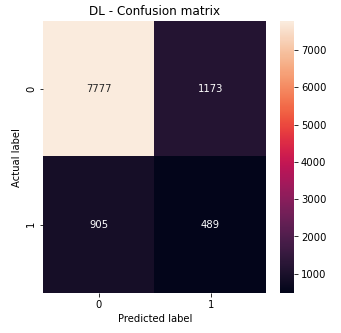}

\end{figure}   
\unskip

\begin{table}[H]
\caption{Hyperparameter tuning (and class weights balancing) - Sklearn GridSearchCV.\label{opt_par1}}
\resizebox{\textwidth}{!}{%
\begin{tabular}{|c|c|c|}
\hline
 &
  \textbf{Parameters Grid} &
  \textbf{Best Parameters} \\ \hline
\textbf{Logistic Regression} &
  \begin{tabular}[c]{@{}c@{}}'max\_iter': {[}20, 50, 100, 200,   500, 1000{]},\\      'solver': {[}'newton-cg', 'lbfgs', 'liblinear', 'sag', 'saga'{]},\\      'class\_weight': {[}'balanced', None{]}\end{tabular} &
  \begin{tabular}[c]{@{}c@{}}'max\_iter': 20\\      'solver': 'sag'\\      'class\_weight': 'balanced'\end{tabular} \\ \hline
\textbf{Decision Tree} &
  \begin{tabular}[c]{@{}c@{}}'criterion': {[}'gini', 'entropy'{]},   \\      'max\_depth': {[}5, 15, 25, None{]}, \\      'max\_features': {[}'auto', 'log2', None{]}, \\      'min\_samples\_split': {[}2, 9, 16{]}, \\      'min\_samples\_leaf': {[}1, 8, 15{]}, \\      'class\_weight': {[}'balanced', None{]}\end{tabular} &
  \begin{tabular}[c]{@{}c@{}}'criterion': 'gini', \\      'max\_depth': None, \\      'max\_features': 'auto', \\      'min\_samples\_split': 2, \\      'min\_samples\_leaf': 1, \\      'class\_weight': 'balanced'\end{tabular} \\ \hline
\textbf{Random Forest} &
  \begin{tabular}[c]{@{}c@{}}'criterion': {[}'gini', 'entropy'{]},   \\      'n\_estimators': {[}100, 200, 300{]}, \\      'max\_depth': {[}5, 15, 25, None{]}, \\      'max\_features': {[}'auto', 'log2', None{]}, \\      'class\_weight': {[}'balanced', 'balanced\_subsample', None{]}\end{tabular} &
  \begin{tabular}[c]{@{}c@{}}'criterion': 'gini', \\      'n\_estimators': 300, \\      'max\_depth': None, \\      'max\_features': 'log2', \\      'class\_weight': 'balanced'\end{tabular} \\ \hline
\textbf{XGBoost} &
  \begin{tabular}[c]{@{}c@{}}'n\_estimators': {[}100, 200, 300{]},   \\      'max\_depth': {[}6, 12, 18, 20{]}, \\      'learning\_rate': {[}0.1, 0.5, 0.9, 0.95{]}, \\      'subsample': {[}0.6, 1.0{]}, \\      'scale\_pos\_weight': {[}1, 6.4095{]}\end{tabular} &
  \begin{tabular}[c]{@{}c@{}}'n\_estimators': 100, \\      'max\_depth': 20, \\      'learning\_rate': 0.5, \\      'subsample': 1.0, \\      'scale\_pos\_weight': 6.4095\end{tabular} \\ \hline
\end{tabular}%
}
%\label{tab:my-table}
\end{table}

\begin{table}[H]
\caption{Hyperparameter tuning (and class weights balancing) - Keras Tuner Hyperband.\label{opt_par2}}
\resizebox{\textwidth}{!}{%
\begin{tabular}{|c|c|c|}
\hline
 &
  \textbf{Parameters Grid} &
  \textbf{Best Parameters} \\ \hline
\textbf{DL} &
  \begin{tabular}[c]{@{}c@{}}'batch\_size': {[}128, 256,   512{]}\\      'num\_layers': 1 - 3\\      'units\_1': 16 - 256, step 16\\      'units\_2': 16 - 256, step 16\\      'units\_3': 16 - 256, step 16\\      'learning\_rate': {[}0.1, 0.01, 0.001, 1e-4, 1e-5, 1e-6{]}\\      'class\_weight': {[}None, \{0: 0.5780, 1: 3,7047\}{]}\end{tabular} &
  \begin{tabular}[c]{@{}c@{}}batch\_size': 256,\\      'num\_layers': 3,\\      'units\_1': 192,\\      'units\_2': 192,\\      'units\_': 240,\\      'learning\_rate': 0.001,\\      'class\_weight': \{0: 0.5780, 1: 3,7047\}\end{tabular} \\ \hline
\end{tabular}%
}
%\label{tab:my-table}
\end{table}

\begin{table}[H] 
\caption{Results of ML models after hyperparameter tuning and class weights balancing.\label{tab3}}
\newcolumntype{C}{>{\centering\arraybackslash}X}
\begin{tabularx}{\textwidth}{CCCCCCC}
\toprule
 & \textbf{accuracy} & \textbf{F1} & \textbf{precision} & \textbf{recall} & \textbf{AUC} & \textbf{AUPRC}\\
\midrule
\textbf{Logistic Regression} & 0.5218 & 0.2421 & 0.1539 & 0.5667 & 0.5407 & 0.1628 \\ 
\textbf{Decision Tree}       & 0.8124 & 0.3418 & 0.3241 & 0.3615 & 0.6221 & 0.215  \\ 
\textbf{Random Forest}       & 0.8498 & 0.3547 & 0.4211 & 0.3063 & 0.6204 & 0.2784 \\ 
\textbf{XGBoost}             & 0.8422 & 0.3549 & 0.3952 & 0.3221 & 0.6227 & 0.275  \\ 
\textbf{DL}                  & 0.7828 & 0.3451 & 0.2906 & 0.4247 & 0.6439 & 0.2629\\ 
\bottomrule
\end{tabularx}
\end{table}
\unskip

\begin{figure}[H]
\caption{Confusion matrix of ML models after SMOTE and hyperparameter tuning and class weight balancing.\label{conf_matrix_SMOTE2}}
%\column{0.33\textwidth}
\centering
\includegraphics[scale=0.65]{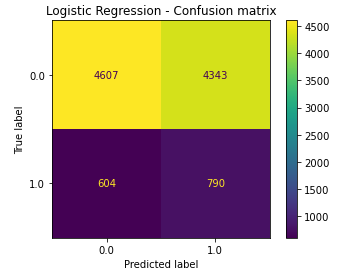}
\includegraphics[scale=0.65]{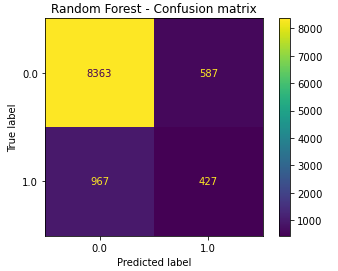}
%\column{0.33\textwidth}
\centering
\includegraphics[scale=0.65]{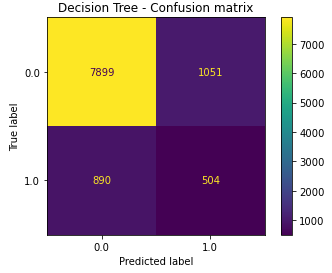}
\includegraphics[scale=0.65]{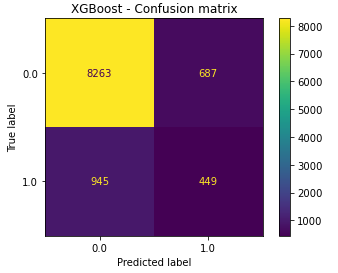}
%\column{0.33\textwidth}
\centering
\includegraphics[scale=0.65]{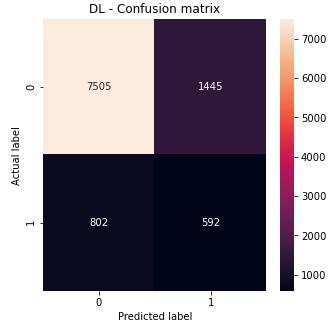}

\end{figure}
\unskip

\begin{figure}[H]
\caption{Decision tree - feature importance \label{decision_tree_FI}}
%\column{0.67\textwidth}
\centering
\includegraphics[scale=0.4]{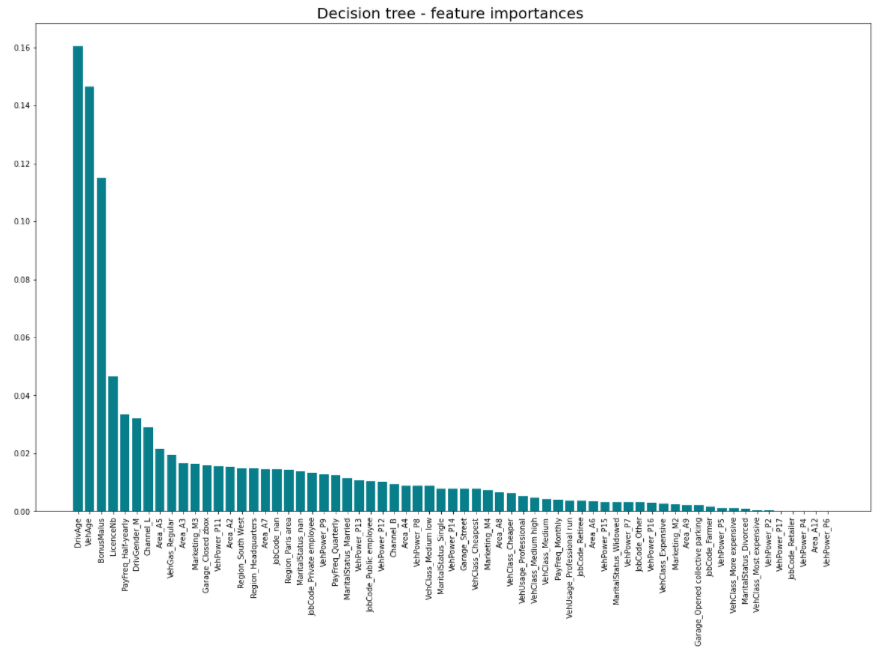}
%\column{0.33\textwidth}
\centering
\includegraphics[scale=0.65]{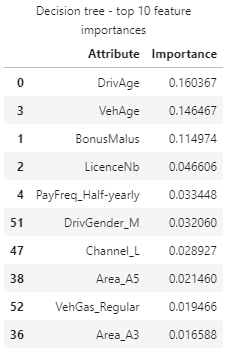}

\end{figure}
\unskip

\begin{figure}[H]
\caption{Random forest - feature importance \label{random_forest_FI}}
%\column{0.67\textwidth}
\centering
\includegraphics[scale=0.4]{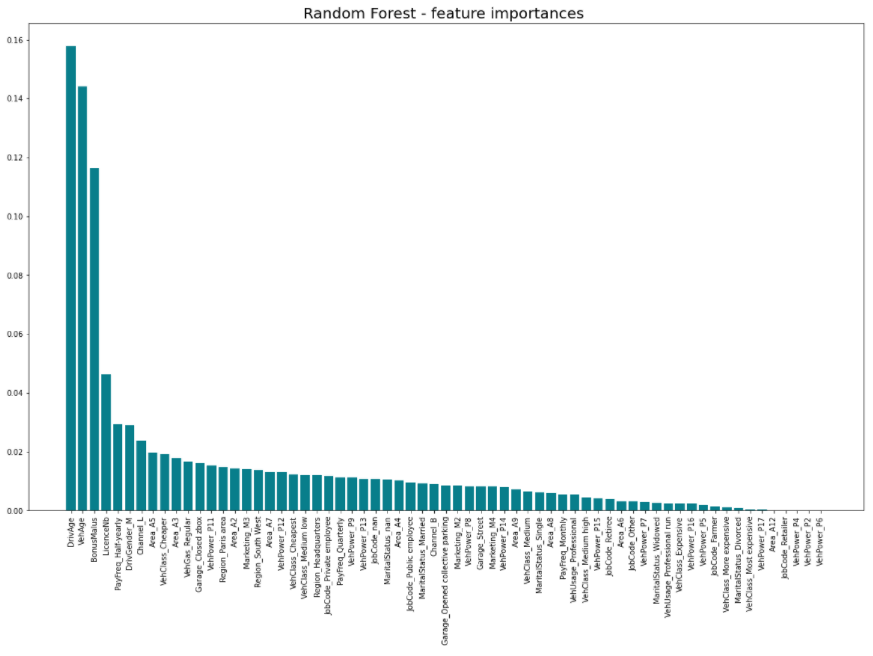}
%\column{0.33\textwidth}
\centering
\includegraphics[scale=0.65]{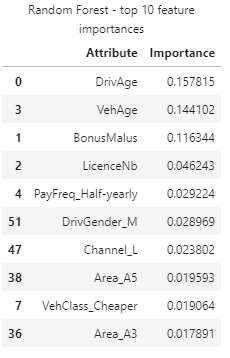}

\end{figure}
\unskip

\begin{figure}[H]
\caption{XGBoost - feature importance \label{xgboost_FI}}
%\column{0.67\textwidth}
\centering
\includegraphics[scale=0.4]{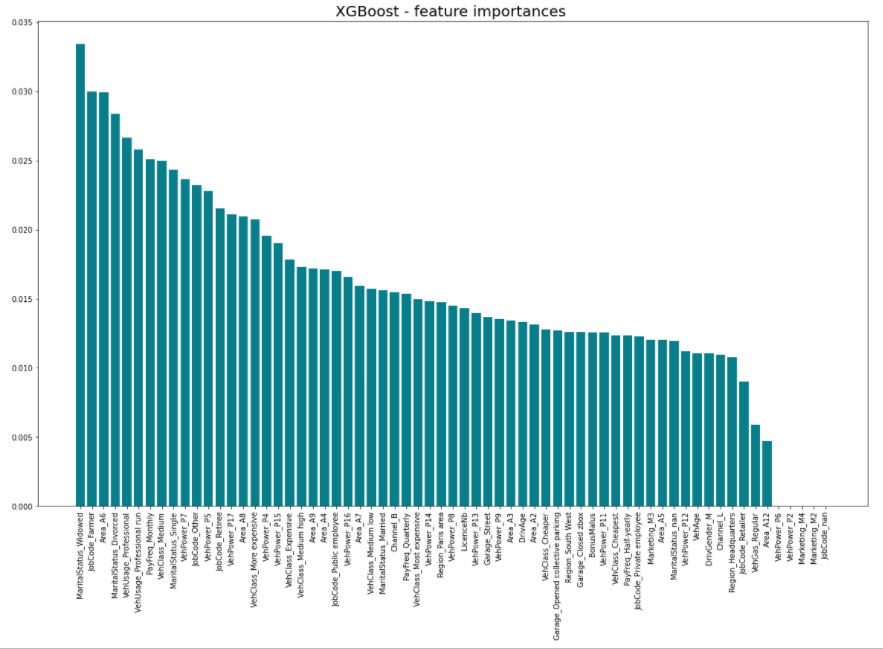}
%\column{0.33\textwidth}
\centering
\includegraphics[scale=0.6]{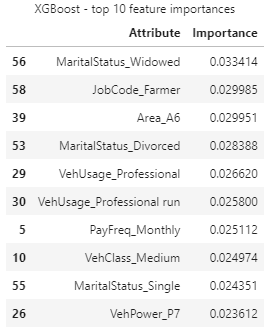}

\end{figure}
\unskip

\end{document}